\documentclass[aps, prl, reprint, notitlepage,superscriptaddress]{revtex4-1}
\usepackage{amsmath}
\usepackage{amssymb}
\usepackage{graphicx}
\usepackage{import}
\usepackage[caption = false]{subfig}
\usepackage{import, xcolor}
\usepackage[retainorgcmds]{IEEEtrantools}
\usepackage{bm}
\usepackage[none]{hyphenat}
\renewcommand{\vec}[1]{\boldsymbol{#1}} % bold symbol for vectors

\usepackage{textcomp}

\newcommand{\ii}{i}
\newcommand{\ee}{\text{e}}

\usepackage[colorlinks,linkcolor=blue,citecolor=blue,urlcolor=blue]{hyperref}

\begin{document}

\title{Quantum, Nonlocal Aberration Cancellation}
\author{A. Nicholas Black}
\affiliation{Department of Physics and Astronomy, University of Rochester, Rochester, NY, USA 14627}
\author{Enno Giese}
\affiliation{Department of Physics, University of Ottawa, Ottawa, ON, Canada K1N 6N5}
\affiliation{Current address: Institut f{\"u}r Quantenphysik and Center for Integrated Quantum
    Science and Technology (IQ\textsuperscript{ST}), Universit{\"a}t Ulm, Albert-Einstein-Allee 11, Ulm, Germany D-89069}
\author{Boris Braverman}
\affiliation{Department of Physics, University of Ottawa, Ottawa, ON, Canada K1N 6N5}
\author{Nicholas Zollo}
\affiliation{The College of Optics and Photonics, University of Central Florida, Orlando, FL, USA 32816}
\author{Stephen M. Barnett}
\affiliation{School of Physics and Astronomy, University of Glasgow, Glasgow, UK G12 8QQ}
\author{Robert W. Boyd}
\affiliation{Department of Physics and Astronomy, University of Rochester, Rochester, NY, USA 14627}
\affiliation{Department of Physics, University of Ottawa, Ottawa, ON, Canada K1N 6N5}
\begin{abstract}
    Phase distortions, or aberrations, can negatively influence the performance of an optical imaging system.  Through the use of position-momentum entangled photons, we nonlocally correct for aberrations in one photon's optical path by intentionally introducing the complementary aberrations in the optical path of the other photon.  In particular, we demonstrate the simultaneous nonlocal cancellation of aberrations that are of both even and odd order in the photons' transverse degrees of freedom.  We also demonstrate a potential application of this technique by nonlocally cancelling the effect of defocus in a quantum imaging experiment and thereby recover the original spatial resolution.
\end{abstract}
\maketitle

%The strong correlations inherent to entangled photons are of interest not only for the foundations of quantum physics~\cite{ ClauserPRL69,AspectPRL82,KwiatPRL95,MairNat01,HowellPRL04,GiustinaPRL15,ShalmPRL15,HandsteinerPRL17,MacLeanPRL18}, but are also routinely used for quantum information processing~\cite{WaltherNat05,ChenPRL07,QiangNatPh18} and quantum imaging~\cite{StrekalovPRL95,PittmanPRA95,Kolobov06,LemosNat14}.
Entangled photons display strong correlations in conjugate continuous variables and are therefore of interest not only for the foundations of quantum physics~\cite{ ClauserPRL69,AspectPRL82,KwiatPRL95,MairNat01,HowellPRL04,GiustinaPRL15,ShalmPRL15,HandsteinerPRL17,MacLeanPRL18}, but are also routinely used for quantum information processing~\cite{WaltherNat05,ChenPRL07,QiangNatPh18} and quantum imaging~\cite{StrekalovPRL95,PittmanPRA95,Kolobov06,LemosNat14}.
Group velocity dispersion (GVD)~\cite{FransonPRA92} and wavefront aberrations~\cite{AbouraddyPRL01} can negatively influence the correlations and entanglement of continuous variables.
In this Letter, we show that through the use of position-momentum entangled photon pairs aberrations experienced by one photon can in a sense be undone by tailoring the wavefront structure of the other photon without ever bringing the two photons back together.
In this sense, we demonstrate \emph{nonlocal} aberration cancellation.
%In this Letter, we show that through the use of position-momentum entangled photon pairs we are able to perform aberration cancellation in a nonlocal manner. More specifically, we show that aberrations experienced by one photon can in a sense be undone by tailoring the wavefront structure of the other photon without ever bringing the two photons back together.

The mitigation of GVD has been proposed and demonstrated both in a local scheme based on indistinguishability~\cite{SteinbergPRL92,JeffersPRA93} and in a nonlocal manner that relied on frequency correlations of the light~\cite{BaekOptEx09, ZhongPRA13, MacLeanPRL18}. The simultaneous~\cite{LukensPRL14, RyuOptEx17} and non-simultaneous~\cite{MinaevaPRL09} cancellation of even- and odd-order terms in the dispersion relation have also been demonstrated in a local manner.  Because frequency correlations do not necessarily have to be of quantum origin, even classical dispersion cancellation techniques have been realized~\cite{TorresCompanyNJP09,ShapiroPRA10, PrevedelPRA11, Torres-CompanyPRL12}. As the spatial analog to dispersion~\cite{KolnerJQE94}, aberrations caused by transverse momentum-dependent phase shifts can be canceled in a conceptually similar manner.
Only local aberration cancellation has been performed experimentally~\cite{BonatoPRL08,FilpiOptEx15}, except for the nonlocal compensation of pure phase objects revealed through polarization correlations~\cite{CialdiPRA11}.
%Except for the nonlocal compensation of pure phase objects revealed through polarization correlations~\cite{CialdiPRA11}, only local aberration cancellation has been performed experimentally for either even-order~\cite{BonatoPRL08} or odd-order~\cite{FilpiOptEx15} aberrations.

We complement these studies by demonstrating both even- and odd-order nonlocal aberration cancellation simultaneously with entangled photon pairs.
Because the observation of position-momentum entanglement is strongly influenced by the propagation of the photons~\cite{ChanPRA07,TascaPRA09}, aberrations in the path of one photon affect the observed entanglement significantly.  In particular, quadratic aberrations that act as defocus lead to a form of entanglement migration~\cite{TascaPRA09}.  These deleterious effects can be cancelled by acting on the state of the second photon with an appropriately chosen aberration.
In contrast to Ref.~\cite{TascaPRA09}, we perform higher-order as well as quadratic aberration cancellation and enhance the quality of quantum imaging in the presence of aberrations~\cite{SimonJOSAB11}.  Our scheme can be extended to the case when the two photons are different frequencies or to implement the spatial analog of the encoding scheme in Ref.~\cite{LukensPRL14}.
\begin{figure}

	\includegraphics[width = \columnwidth]{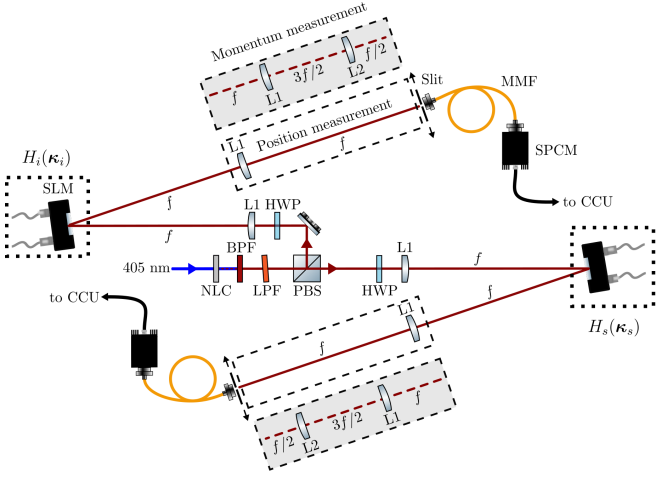}
	\caption{Experimental setup for performing aberration cancellation using position-momentum entangled photon pairs.  The signal and idler photons created in nearly collinear degenerate type-II spontaneous parametric down-conversion are split by a polarizing beamsplitter (PBS) and pass through $4f$ imaging systems for the position-basis coincidence measurement.  Spatially-resolved detection is achieved using $100$ \textmu m wide translatable slits.  Spatial light modulators (SLM) located in the Fourier plane of the crystal introduce aberrations, $H_{j}(\vec{\kappa}_{j}),\;j = s,i$, to each path.  For the momentum-basis coincidence measurement, the SLMs are imaged onto the plane of the slits.  The lenses L1 and L2 have focal lengths $f = 40\,\mathrm{cm}$ and $f$/2 respectively.  (BPF--band-pass filter; CCU--coincidence counting unit; HWP--half-wave plate; LPF--low-pass filter; MMF--multimode fiber; NLC--nonlinear crystal; SPCM--single-photon counting module)
	}
	\label{fig:expsetup}
\end{figure}

Figure~\ref{fig:expsetup} shows the experimental setup (explained later in more detail) whereby aberrations can be introduced controllably to a photon pair created in nearly collinear type-II spontaneous parametric down-conversion (SPDC).
The two photons, called signal and idler, are entangled in their transverse degrees of freedom and can be described by the joint wave function in the momentum representation after passing through the optical system in Fig.~\ref{fig:expsetup},
\begin{equation}
    \label{eq:psi_momentum}
    \psi(\vec{\kappa}_s,\vec{\kappa}_i) = C \mathcal{E}(\vec{\kappa}_s+\vec{\kappa}_i)\tilde{\chi}^{(2)}(\Delta k_z)H_s(\vec{\kappa}_s)H_{i}(\vec{\kappa}_i),
\end{equation}
where $\vec{\kappa}_s$ and $\vec{\kappa}_i$ denote the transverse components of the wave vectors of the signal and idler, respectively.  The constant $C$ includes terms resulting from the quantization of the electric field and the nonlinear interaction \footnote{The proportionality constant in Eq.~\eqref{eq:psi_momentum} for the special case of box-shaped nonlinear crystal is
\begin{equation}
C = \frac{\ii V_{Q} \ell \chi^{(2)} \omega_{p}}{2^{5} \pi^{3} c v_{p}(\omega_{p})\sqrt{n_{o}(\omega_{p}/2)n_{e}(\omega_{p}/2)v_{s}(\omega_{p}/2)v_{i}(\omega_{p}/2)}}.
\end{equation}
Here, $V_{Q}$ is the quantization volume, $\ell$ is length of nonlinear crystal, and $\chi^{(2)}$ is the second-order nonlinear susceptibility for the type-II interaction. The frequency $\omega_{p}$ is the angular frequency of the pump and $n_{\sigma}(\omega_{p}/2)$, where $\sigma = e,o$, is the extraordinary or ordinary index of refraction respectively. The interaction time has been assumed to be long enough so that the time between down-conversion events is longer than the resolving time of the detectors so that $\omega_s + \omega_i $ corresponds to the frequency of the pump.  The signal and idler frequencies are assumed to be well-defined by filters around $\omega_{p}/2$. $v_{j}$, where $j = p,s,i$, is the group velocity of the pump, signal, or idler respectively.}. The angular profile of the pump beam, $\mathcal{E}$, controls the anticorrelation of the signal and idler momenta \cite{JhaPRA10, GiesePhysScrpt18}.  For a crystal whose transverse dimensions are large compared with the pump beam diameter, $\mathcal{E}$ is a function of $\vec{\kappa}_s+\vec{\kappa}_i = \vec{\kappa}_p$, an expression of momentum conservation.  Perfect anticorrelation in the signal and idler momenta is obtained when the pump is a plane wave, $\mathcal{E}(\vec{\kappa}_s+\vec{\kappa}_i) \propto \delta(\vec{\kappa}_s+\vec{\kappa}_i)$.  However, this is generally not the case.

For degenerate SPDC in the paraxial approximation and neglecting walk-off, the longitudinal wave vector mismatch $\Delta k_z$ of pump, signal, and idler fields reduces to $\Delta k_z\cong (\vec{\kappa}_s-\vec{\kappa}_i)^2/(2k_p)$, where $k_p$ is the wave vector of the pump. The wave vector mismatch is engaged through the phase-matching function $\tilde \chi^{(2)}(\Delta k_z)$, which is the Fourier transformation of the longitudinal profile of the nonlinearity.  In the case of a uniform crystal of length $\ell$, it takes the form
\begin{equation}
    \label{eq:phasematcheq}
    \tilde{\chi}^{(2)}(\Delta k_{z})\propto \!\ee^{-\ii\frac{\ell\Delta k_z }{2} } \operatorname{sinc} \!\frac{\ell\Delta k_z }{2} \cong\!\ee^{-\frac{\ell (\vec{\kappa}_s-\vec{\kappa}_i)^2}{4 k_p}(\alpha + \ii)}. \nonumber
\end{equation}
In the last step the $\operatorname{sinc}$ was approximated by a Gaussian with $\alpha= 0.455$~\cite{ChanPRA07}.

The imaging system amplitude transfer functions, $H_j(\vec{\kappa}_j)$ with $j=s,i$, in Eq.~\eqref{eq:psi_momentum} describe the propagation along the signal and idler path, respectively and can also be used to describe quantum ghost imaging~\cite{AbouraddyPRL01}.
For the aberrations considered here, they reduce to a momentum-dependent phase factor $H_j(\vec{\kappa}_j)= \exp[\ii \phi_j(\vec{\kappa}_j)]$.  

\begin{figure*}

	\includegraphics[width = 0.8\textwidth]{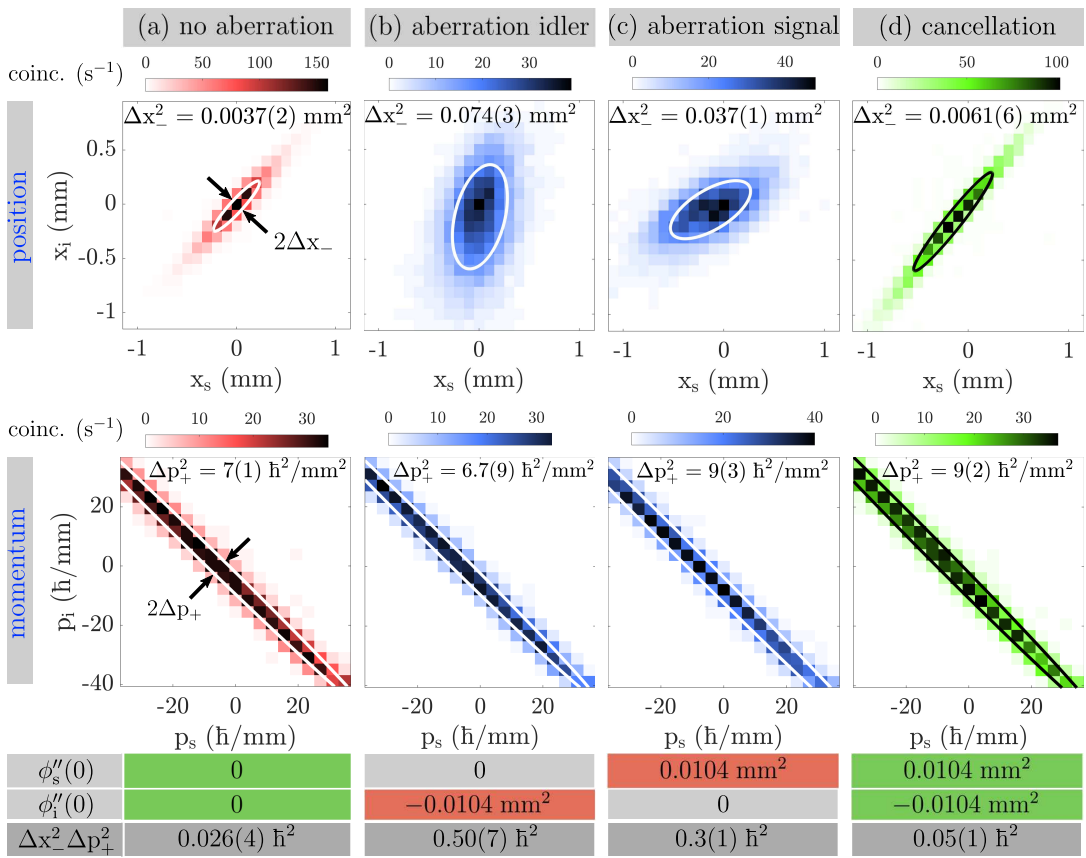}

	\caption{
	Coincidence distributions in the positions (top) and momenta (bottom) of signal and idler photons with (a)~no aberrations, (b,c)~quadratic aberrations introduced solely on the idler or signal, respectively, and (d)~aberration cancellation. The coefficients of the quadratic aberrations are listed below the figure. The momentum distributions remain nearly unaffected by aberrations, whereas the position distributions change significantly. This broadening can be compensated by introducing quadratic aberration to the signal path that is equal in magnitude but opposite in sign from quadratic aberration in the idler path. The white (a,b,c) or black (d) ellipses mark the $1\sigma$ levels of Gaussian fits and are used to obtain $\Delta x_-$ and $\Delta p_+$, necessary for the evaluation of the entanglement criterion (bottom row).
		}
	\label{fig:entangmeas}
\end{figure*}

In the experiment, aberrations are introduced in only one transverse dimension.
For this reason, a one-dimensional version of Eq.~\eqref{eq:psi_momentum} will be considered and the momentum will be defined through the de Broglie relation $p_j = \hbar \kappa_j$ in this particular direction.
The value of $\kappa_j$ can be inferred in the Fourier plane of the crystal from the measured position $\varrho_j = f \kappa_j/k_j$, where $f$ is the focal length of the lens used to access the Fourier plane.

The joint momentum distribution is given by the modulus squared of Eq.~\eqref{eq:psi_momentum},
\begin{equation}
    \label{eq:momprobdist}
    P(p_s,p_i) =   |C|^2|\mathcal{E}(\kappa_s+\kappa_i)|^2 |\tilde{\chi}^{(2)}(\Delta \kappa)|^2 .
\end{equation}
Equation~\eqref{eq:momprobdist} can be written approximately as a two-dimensional Gaussian whose variance, $\Delta p_{+}^{2}$, in direction $p_+=(p_s+p_i)/\sqrt{2}$ is determined by the angular profile of the pump. In direction $p_-=(p_s-p_i)/\sqrt{2}$ the variance is determined by crystal properties, $\Delta p_-^2 = \hbar^2 k_p/(\alpha \ell)$.  Because the transfer functions for aberrations, $H_{j}$, are a multiplicative phase in the momentum basis, they do not appear in Eq.~\eqref{eq:momprobdist}.
%The coincidence measurements in the second row of Fig.~\ref{fig:entangmeas} demonstrate that this description is a good approximation and that aberrations do not have a significant effect on the joint momentum distribution.

In contrast, phases caused by aberrations have significant impact on the joint position distribution, which can be obtained from the Fourier transformation of $\psi(\kappa_s,\kappa_i)$.  Expanding the phase of $H_{s}(\kappa_{s})H_{i}(\kappa_{i})$ in a Taylor series and assuming a plane-wave pump such that $\kappa_{s} = -\kappa_{i} = \kappa$, one finds,
\begin{equation}
    \label{eq:phaseexpansion}
\begin{split} 
    \phi_{s}(\kappa) + \phi_{i}(-\kappa)=&\phi_{s}(0) + \phi_{i}(0)\\
    &+\kappa\left[\phi'_{s}(0) - \phi'_{i}(0)\right]\\ 
    &+\kappa^{2}\left[\phi''_{s}(0) + \phi''_{i}(0)\right]/2!\\ 
    &+\kappa^{3}\left[\phi'''_{s}(0) - \phi'''_{i}(0)\right]/3!+ \dots 
\end{split}
\end{equation}
By the Fourier shift theorem, the first-order term in the transfer functions' phases shifts center of the joint position distribution.  The second-order term in Eq.~\eqref{eq:phaseexpansion} changes the variance of the joint position distribution since the Fourier transform of a Gaussian is also Gaussian.  Analogously, the third-order term introduces skew to the joint position distribution.  From Eq.~\eqref{eq:phaseexpansion} it is clear that when the even-order derivatives of the individual phase functions are equal in magnitude, but opposite in sign, and the odd-order derivatives are equal, the effects of aberrations can be cancelled nonlocally.  This leads to the condition for all-order aberration cancellation under the plane-wave-pump approximation, $\phi_{s}(\kappa) = -\phi_{i}(-\kappa)$.  If the pump is not a plane wave, the momenta of the signal and idler photons are no longer perfectly anticorrelated, making complete aberration cancellation impossible.

Position- and momentum-correlation measurements were carried out to study the effects of introducing aberrations to the signal and idler beams separately.
Figure~\ref{fig:expsetup} shows the experimental setup where a $\beta$-barium borate (BBO) crystal was pumped under nearly collinear type-II phase matching by an $\sim19\,\mathrm{mW}$ collimated Gaussian beam with a diameter of $\sim1\,\mathrm{mm}$ centered at $405\,\mathrm{nm}$ to create transversely entangled photons.
The output signal and idler photons were spectrally filtered with a narrowband ($10\,\mathrm{nm}$) filter centered at $810\,\mathrm{nm}$ followed by a long-pass filter with the cutoff wavelength at $594\,\mathrm{nm}$.
Spatial light modulators (SLM) placed in the Fourier plane of the crystal along the signal and idler paths implemented the transverse momentum-dependent phase shifts leading to aberration.  Due to alignment-related quadratic aberrations (defocus) in the experimental setup, an additional defocus with $\phi_{s}''(0) = -0.0052\,\mathrm{mm^{2}}$ was introduced to the signal path for all measurements.
Slits of width $100\,$\textmu m were used to select a small portion of the beam for detection and were translated with servo-controlled micrometer stages in steps of $100$\,\textmu m.
For each setting of the two slits, the coincidence detection rate was recorded using single photon detectors and a coincidence-counting unit~\cite{BranningAJP09} with a coincidence window of $\sim 13\,\mathrm{ns}$.  The SLMs were imaged onto the slit plane to detect the momentum distribution (Fig.~\ref{fig:entangmeas}, second row) and brought to the far field to detect the position distribution (Fig.~\ref{fig:entangmeas}, first row).

The results of introducing quadratic aberrations are shown in Fig.~\ref{fig:entangmeas}, which displays the coincidence counts in both the position and momentum representations for the case of no aberrations (a), for aberrations on idler only (b) and on signal only (c), and for the cancellation scheme (d).
As predicted, the momentum distributions (bottom) do not vary significantly.
However, the widths of the position distributions (top) broaden as soon as aberrations are introduced.
The broadening is a direct consequence of the defocus in one arm, which causes a broadening of the respective marginal probability distribution.
Notably, broadening along the anti-diagonal direction $x_{-} = (x_{s} - x_{i})/\sqrt{2}$ of $(x_{s}, x_{i})$-space is almost completely cancelled by introducing opposite aberrations to both branches, as seen in Fig.~\ref{fig:entangmeas}(d).

Figure~\ref{fig:entangmeas} shows that the correlation between the positions of the photons depends on aberrations, which in turn also affect the Heisenberg-type inequality
\begin{equation}
\label{eq:Mancini}
\Delta x_{-}^{2}\Delta p_{+}^{2} \geq \hbar^{2}/4
\end{equation}
whose violation is commonly used to verify position-momentum entanglement~\cite{ReidPRL88,ManciniPRL02,HowellPRL04,GiesePhysScrpt18}.  Violating the inequality from Eq.~\eqref{eq:Mancini} is a signature for entanglement.
To determine the widths $\Delta x_{-}$ and $\Delta p_{+}$ from the experimental coincidence distribution in Fig.~\ref{fig:entangmeas}, we perform maximum likelihood fitting using the model of a bivariate Gaussian distribution and obtain their errors through a Monte Carlo simulation.

\begin{figure}

	\includegraphics[width = \columnwidth]{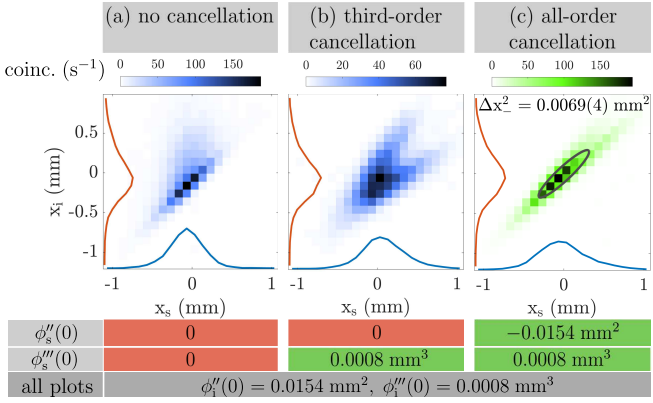}

	\caption{
	Higher-order aberration cancellation. (a)~Second- and third-order aberrations are introduced on the idler (coefficients on the bottom) leading to the displayed coincidence distribution in the position basis. (b)~When third-order aberrations are cancelled, the distribution broadens due to the non-cancelled quadratic aberration. The forked structure is due to the asymmetry of the signal and idler marginal distributions. (c)~Cancellation of all orders.  Inset are the signal and idler marginal distributions (blue and orange lines, respectively).}
	\label{fig:evenodd}
\end{figure}

The table below Fig.~\ref{fig:entangmeas}(a) shows that the generated photon pair is position-momentum entangled without aberrations present since the data violate the inequality from Eq.~\eqref{eq:Mancini}.
However, introducing second-order aberrations on either the idler or the signal increases the variance $\Delta x_{-}^2$.  Therefore, Eq.~\eqref{eq:Mancini} is fulfilled, meaning that entanglement cannot be verified with aberrations present.
It must be emphasized that measurements satisfying Eq.~\eqref{eq:Mancini} do not imply a lack of entanglement since correlations can also exist in the phase of the wave function.
In fact, quadratic aberrations act as a defocus and the two-photon propagation is otherwise unitary so that we observe a version of entanglement migration~\cite{ChanPRA07}.  

After nonlocal aberration cancellation is performed by choosing $\phi''_s(0)= - \phi''_i(0)$, the variance $\Delta x_-^2$ decreases significantly and entanglement is verified.
In this sense, the effect of entanglement migration can be undone.
However, the values displayed in Fig.~\ref{fig:entangmeas}(d) show that the product $\Delta x_-^2 \Delta p_+^2$ is still larger than for the case without aberrations.

As already suggested by the discussion above, the cancellation is not perfect because the pump beam is not a plane wave, resulting in imperfect anti-correlation of the signal and idler momenta.
In this case, the exact Fourier transformation of $\psi(\kappa_s,\kappa_i)$ with a Gaussian beam profile $\mathcal{E}(\kappa_{s} + \kappa_{i})$ of width $\Delta \kappa_p$ and within the Gaussian approximation of the phase-matching function gives the probability distribution $ P(x_{s}, x_{i}) \propto \exp[- x_-^2/ (2 \Delta x_-^2)]$ with $x_s = -x_i$, which is of Gaussian form along the anti-diagonal $x_-$.  Even under the condition for aberration cancellation $\phi''_s(0)= - \phi''_i(0)= \beta$, the variance
\begin{equation}\label{eq:Delta_x_-}
    \Delta x_-^2 = \frac{[\ell \alpha + k_p \beta^2 \Delta \kappa_p^2]^2 + \ell^2}{2 k_p \ell \alpha + 2 k_p^2 \beta^2 \Delta \kappa_p^2} 
\end{equation}
still depends on $\beta$ due to the finite size of the pump profile $\Delta \kappa_p$,  and perfect aberration cancellation cannot be achieved.
To demonstrate the effect explicitly, one can expand Eq.~\eqref{eq:Delta_x_-} up to second order in $\Delta \kappa_p$
\begin{equation}
\label{eq:width_expansion}
    \Delta x_-^2 \cong \frac{\ell (\alpha^2+1)}{2 k_p \alpha} + \frac{ 1-\alpha^2}{2  \alpha^2} \beta^2 \Delta \kappa_p^2 
\end{equation}
and see that the zeroth-order term corresponds to the width without aberrations, and $\Delta x_{-}^2$ will increase for any finite $\beta$ and $\Delta \kappa_{p}$.  Non-cancelled aberrations also lead to a rotation of the joint position distribution, as seen in Figs.~\ref{fig:entangmeas}(b) and~\ref{fig:entangmeas}(c), due to the appearance of a correlation term in $P(x_{+},x_{-})$.

The impact of the pump profile is even more significant when canceling higher-order aberrations.
Figure~\ref{fig:evenodd}(a) shows the coincidence distribution in the position representation when both quadratic and cubic aberrations are introduced into the idler path only.
Compared with Fig.~\ref{fig:entangmeas}(b), the broadened joint distribution is clearly skewed by the introduction of cubic aberration.
According to the cancellation scheme discussed previously, the coefficients of the cubic phase terms in each path are chosen to be the same, Fig.~\ref{fig:evenodd}(b).  The forked structure in Fig.~\ref{fig:evenodd}(b) is a result of the fact that the marginal probability distributions of the signal and idler must be skewed, but only the idler's marginal distribution can be broadened.
When both quadratic and cubic aberration cancellation is performed, Fig.~\ref{fig:evenodd}(c), the distribution approaches the non-aberrated distribution, Fig.~\ref{fig:entangmeas}(a), but displays an asymmetry along the $x_{+}$ direction.  Higher-order aberration cancellation appears to be more sensitive to the finite width of the pump profile because the total amount of aberrations introduced is larger than for the quadratic scheme.

\begin{figure}

	\includegraphics[width = \columnwidth]{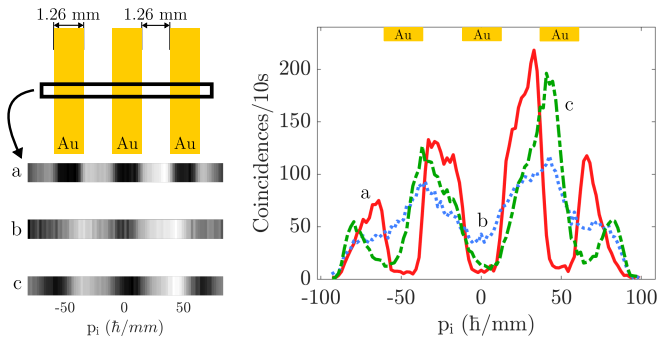}

	\caption{(left) The top panel shows the  horizontal width of the bars and the bottom panel shows psuedo-2D images of the gold bars with saturation for each scenario on the right. (right) The solid red line (a) is the coincidence image of three Au bars placed in the signal path momentum plane without any aberrations present.  The dotted blue line (b) is the coincidence image with phase aberrations in the idler path only, according to $\theta_{i}(x_{i}) = \theta_{i}''(0) x_{i}^{2}/2$, where $\theta_{i}''(0) = 73.7\;\mathrm{mm^{-2}}$.  When the defocus is cancelled nonlocally, $\theta_{s}''(0) = -\theta_{i}''(0)$, the coincidence image of the slits is almost completely recovered (dashed green line, c).}
	\label{fig:quantimage}
\end{figure}

To demonstrate the utility of nonlocal aberration cancellation, quantum imaging in the presence of focusing error and its nonlocal cancellation was performed in one dimension. Three parallel gold bars placed in front of a bucket detector in the signal path constituted the image.
To implement the bucket detector, photons passing through the gold bars were coupled into a multi-mode fiber with a 400\,\textmu m core diameter (NA = 0.39) using a 10x microscope objective (NA = 0.25) and were detected using the same single photon counting modules as in Fig.~\ref{fig:expsetup}.  Unlike the experiments shown in Fig.~\ref{fig:expsetup}, aberrations were introduced in the image plane of the crystal and coincidence measurements (imaging) took place in the Fourier plane. This configuration was chosen to take advantage of the larger beam cross section in the Fourier plane of the crystal.

From the results shown in Fig.~\ref{fig:quantimage}(right), the image of the slits (solid red line) is clearly lost after the introduction of quadratic aberrations (dotted blue line) and then partially recovered with aberration cancellation (dashed green line), demonstrating the utility of this effect in quantum imaging.  The decrease in contrast and the increased period of the dashed green line compared to the solid red line is due to the finite width of phase-matching function.  In this case, the effect of the phase-matching function is analogous to that of using a Gaussian pump beam when aberrations are introduced in the momentum representation.

In conclusion, we have demonstrated the first nonlocal cancellation of even- and odd-order aberrations simultaneously. Furthermore, we have shown how aberrations and their subsequent nonlocal cancellation influence the results of transverse entanglement measurements using the criterion from Eq.~\eqref{eq:Mancini}. We also applied this technique to nonlocally correct for focusing error in a quantum imaging setup. Prior theoretical and experimental work has shown that both dispersion cancellation \cite{TorresCompanyNJP09,ShapiroPRA10,PrevedelPRA11,Torres-CompanyPRL12} and ghost imaging \cite{BenninkPRL02} are possible using classically correlated light beams.  Such demonstrations suggest that it may also be possible to observe nonlocal aberration cancellation using a light source with classical, rather than quantum mechanical, correlations.
\begin{acknowledgements}
A.N.B. and R.W.B. acknowledge the generous support of the Office of Naval Research under award No. N00014-17-1-2443.  A.N.B. and E.G. would like to thank Robert Fickler for fruitful discussions during the experiment.  B.B. acknowledges the support of the Banting Postdoctoral Fellowship.  S.M.B. thanks the Royal Society for the support of a Research Professorship RP150122.
\end{acknowledgements}

%\bibliography{NonlocalAberrationCancellationNickBlackBib}

\end{document}